\documentclass[superscriptaddress,aps,pra,showpacs,twocolumn,nofootinbib,longbibliography,notitlepage]{revtex4-2}
\usepackage{etex}
\usepackage{amsmath,amssymb,amsthm}
\usepackage[colorlinks=true,citecolor=blue,urlcolor=blue]{hyperref}
\usepackage[pdftex]{graphicx}
\usepackage{times,txfonts}
\usepackage{braket}
\usepackage{color}
\usepackage{natbib}
\usepackage{amsmath,blkarray}
\usepackage{mathtools}
\usepackage{latexsym}
\usepackage{tabularx, booktabs}
\usepackage{graphics,epstopdf}
\usepackage{graphicx}
\usepackage{float}
\usepackage{graphicx}
\usepackage{amsfonts}
\usepackage{subcaption}
\usepackage{color,soul}

\newcommand{\be}{\begin{equation}}
	\newcommand{\ee}{\end{equation}}
\newcommand{\ba}{\end{eqnarray}}
\newcommand{\ea}{\end{eqnarray}}

\begin{document}
\title{Generalized  $n$-locality inequalities in linear-chain network for arbitrary inputs scenario and their quantum violations}
\author{Rahul Kumar}
\author{ A. K. Pan }
\email{akp@nitp.ac.in}
\affiliation{National Institute of Technology Patna, Ashok Rajpath, Patna, Bihar 800005, India}
\begin{abstract}
Multipartite nonlocality in a network is conceptually different from standard multipartite Bell nonlocality. In recent times,  network nonlocality has been studied for various topologies. We consider a  linear-chain topology of the network and demonstrate the quantum nonlocality (the non-$n$-locality). Such a network scenario involves $n$ number of independent sources and $n+1$ parties, two edge parties (Alice and Charlie), and $n-1$ central parties (Bobs). It is commonly assumed that each party receives only two inputs. In this work, we consider a generalized scenario where the edge parties receive an arbitrary  $n$ number of inputs  (equals to a number of independent sources), and each of the central parties receives two inputs. We derive a family of generalized $n$-locality inequalities for a linear-chain network for arbitrary $n$ and demonstrate the optimal quantum violation of the inequalities. We introduce an elegant sum-of-squares approach enabling the derivation of the optimal quantum violation of aforesaid inequalities \emph{without} assuming the dimension of the system. We show that the optimal quantum violation requires the observables of edge parties to mutually anticommuting. For $n=2$ and $3$,  the optimal quantum violation can be obtained when each edge party shares a two-qubit entangled state with central parties. We further argue that for $n\geq 2$, a single copy of a two-qubit-entangled state may not be enough to exhibit the violation of $n$-locality inequality, but multiple copies of it can activate the quantum violation.  
\end{abstract}
\maketitle
Distant observers sharing an entangled state can generate quantum correlations that cannot be reproduced by local realistic theory \cite{bell}. This fundamental feature is commonly known as quantum nonlocality and is the hallmark property of quantum theory. Quantum nonlocality is commonly revealed through the quantum violation of a suitable Bell inequality. Not surprisingly, various forms of Bell inequalities have been derived  \cite{brunnerreview} over the years for bipartite as well as the multipartite scenario. In recent times, there has been an upsurge of interest in studying a different form of multipartite nonlocality in quantum networks. In a standard multiparty Bell experiment, one common source distributes the entangled particles to many spatially separated parties. In contrast, the network Bell experiment features multiple independent sources distributing entangled subsystems to the parties. 

The simplest nontrivial network scenario is the bilocality scenario featuring two independent sources and three distant parties. Such a demonstration of non-bilocal correlation in the tripartite network requires two entangled states originating from two independent sources and appropriate choices of measurements for each party. Branciard \emph{et al.,}\cite{Bran2010,Cyril2012}  first derived an interesting form of nonlinear bilocality inequality, which is violated by quantum theory. It is shown that bilocality inequality is violated by all pure two-qubit entangled quantum states \cite{Gisin2017}.

The nonlocal correlation in the quantum network has been extensively studied. In recent times, for various topologies of quantum network and a  flurry of works has recently been reported \cite{Tavakoli2016,Frit2016,Rosse2016,Chav2016,Gisin2017,Tava2017,Luo2018,Gupta2018,Wolfe2019,Cyril2019,Kerstjens2019,Aberg2020,Banerjee2020,Mukh2016,Tava2016,Tavakoliarxiv,Gisinarxiv2021,Kraftarxiv,Frit2012,Armi2014}. An obvious generalization of bilocal network scenario is the $n$-local network scenario  \cite{Frit2012,Armi2014} involving $n$ independent sources, each shares physical systems to two parties. In such scenario, $n$-locality inequalities have been proposed and quantum violation of them are demonstrated \cite{Armi2014,Andr2017,kundu2020,munshi2021,munshi2021a}. Some of the network inequalities have experimentally been tested \cite{Saunders2017,Carvacho2017,Andreoli2017,Poderini2020}. Many interesting  forms of nonlocal correlations in close network have been studied without using any input \cite{Renou2019,Renou2022,Reno2022} for all parties by only considering the output statistics of fixed measurement input. In \cite{Gisi2020}, the constraint on nonlocal correlations in networks has been explored by assuming only the no-signaling condition and independence of the sources. It is also shown that quantum nonlocality in a network can be revealed by allowing an arbitrarily small level of independence between sources \cite{Supic2020}. The device-independent information processing on quantum networks has also been discussed \cite{Lee2018}. The concept of full network nonlocality is recently introduced \cite{Kerst2022} that demonstrates all links in a network distribute nonlocal resources. Self-testing using a quantum network has also been studied \cite{Supic2022}. Note that almost all the $n$-locality inequalities in the open network scenario can be traced back to some forms of Bell inequalities \cite{Gisin2017,Andr2017,kundu2020,munshi2021,munshi2021a}. In a recent work\cite{Tava2021}, interesting bilocality inequalities are formulated that cannot be traced back to a Bell inequality.

This paper considers the linear-chain quantum network that features $n$ independent sources and $n+1$ observers. Such a network involves two edge parties, Alice and Charlie, and an arbitrary $n-1$ number of central parties, Bobs. Both the edge parties received the physical systems from a single source, but the other central parties received inputs from two nearest sources. Compared to other topologies of a quantum network, the $n$-locality inequalities in the linear-chain network are briefly explored \cite{kundu2020,Andr2017}. It is crucial to note that existing works in linear-chain networks assume that each party receives only two inputs. 

We provide a nontrivial generalization of the linear-chain network by considering that Alice and Charlie receive an arbitrary $n$ number of inputs instead of two inputs in existing works. Each of $n-1$ Bobs receives two inputs who together perform $2^{n-1}$ number of binary outcome joint measurements. We derive a family of generalized $n$-locality inequalities in such a network configuration and demonstrate their optimal quantum violation. We introduce an elegant sum-of-squares (SOS) approach, which enables us a dimension-independent derivation of the optimal quantum value of $n$-locality. We further demonstrate the optimal quantum violation of $n$-locality inequality requires $n$ number of mutually anticommuting observables for Alice and Charlie. Note that, for $n=2$,  the optimal quantum violation of $n$-locality inequality can be achieved for the two-qubit entangled state, which is shared between each edge and central parties is required. Similar condition holds for $n=3$. Further, we demonstrate that the optimal quantum violation of $n$-locality inequalities for arbitrary $n$ \emph{without} assuming the dimension of the system, which is in contrast to the earlier works. We show that for $n=2,3$ two-qubit entangled state shared by the edge party and the central party is enough to get the optimal quantum violation. However, the number of anticommuting observables is restricted for a given dimension of the Hilbert space. We show that for $n>3$, a higher-dimensional system is required to obtain the optimal quantum violation of $n$-locality inequalities. 

The paper is organized as follows. In Sec. I,  we discuss the standard bilocality scenario. In Sec. II, by considering the linear-chain network, we extend it to the trilocal scenario and derive optimal quantum violation of trilocality inequality. In Sec. III, we generalize the linear-chain network for an arbitrary $n$  input scenario for edge parties. We propose a  $n$-locality inequality and derive optimal quantum violation without assuming the dimension of the system. Finally, we discuss our results in Sec. IV.
\section{Bilocality scenario}
\begin{figure}[ht]
\centering
{{\includegraphics[width=1\linewidth]{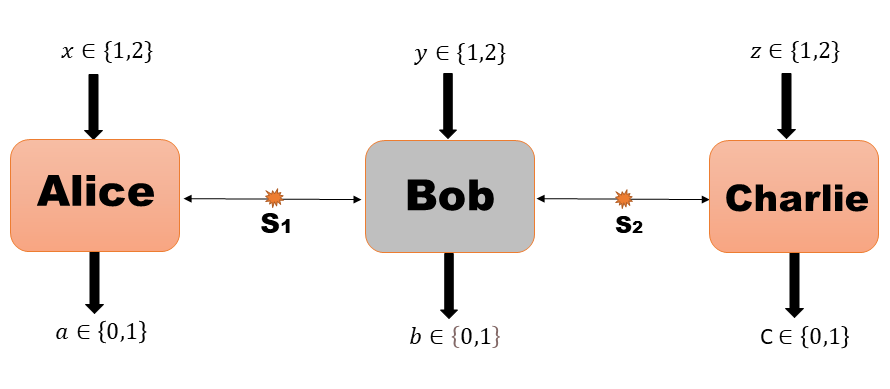}}}
\caption{Bilocal scenario.}
\label{fig:my_label}
\end{figure}
We start by recaptuating the simplest non-trivial network - the well known bilocality scenario \cite{Cyril2012} for ($n=2$), featuring two independent sources $S_1$ and $S_2$ and total three parties Alice, Bob and Charlie. The source $S_{1}$ sends physical systems to Alice and Bob and the source $S_{2}$ sends physical systems to Bob and Charlie. In the standard bilocal scenario $(n=2)$, Alice (Charlie) performs two dichotomic measurements $A_{1}$ and $A_{2}$ ($C_{1}$ and $C_{2}$) corresponding to the input $x\in\{1,2\}$ ($z\in\{1,2\}$), and Bob also performs the measurement of two observables $B_{1}$ and $B_{2}$ corresponding to the input $y\in\{1,2\}$ and produce outputs $a, b, c \in \{0,1\}$ respectively. In particular, Bob performs  measurement on the joint physical systems he receives from two sources $S_{1}$ and $S_{2}$. It is natural to assume that two different states  $\lambda_{1}$ and $\lambda_{2}$ having distribution $\rho(\lambda_{1},\lambda_{2})$ is assigned by the local model of each source where $ \lambda_{1},\lambda_{2}$ are so-called hidden variables. Since $S_{1}$ and $S_{2}$ are assumed to be independent and uncorrelated, $\lambda_{1}$ and $\lambda_{2}$ should be independent and the joint distribution  $\rho{(\lambda_{1},\lambda_{2})}$ can be factorised as  $\rho{(\lambda_{1},\lambda_{2})}$ = $\rho_{1}{(\lambda_{1})}$ $\rho_{2}{(\lambda_{2})}$, and $\rho_{1}{(\lambda_{1})}$ and $\rho_{2}{(\lambda_{2})}$ are independent distribution that satisfy  $\int d\lambda_{1}\rho_{1}{(\lambda_{1})}=1$ and $\int d\lambda_{2}\rho_{2}{(\lambda_{2})}=1$. This constitute the bilocality condition  \cite{Cyril2012}, and the joint probability can be written as
\begin{eqnarray}
	\nonumber
	P(a, b, c|x,y,z)&=&\int\int d\lambda_{1} d\lambda_{2}\hspace{2mm}\rho_{1}(\lambda_{1})\hspace{1mm}\rho_{2}(\lambda_{2})\\
	&&\times P(a|x,\lambda_{1})P(b|y,\lambda_{1},\lambda_{2})P(c|z,\lambda_{3}).\hspace{9pt}
\end{eqnarray}
Alice's and Charlie's outcomes  solely depend on $\lambda_{1}$ and  $\lambda_{2}$ respectively, but Bob's outcome depends on both $\lambda_{1}$ and $\lambda_{2}$.   It is argued in \cite{Cyril2012} that bilocality condition for $P(a,b,c|x,y,z)$ holds if the following nonlinear bilocality inequality 
\begin{equation}
	\label{Delbl22}
	(\beta_{2})_{bl}=\sqrt{|{J}_{2,1}|}+\sqrt{|{J}_{2,2}|}\leq {2}
\end{equation}
is satisfied {where '$bl$' denotes bi-locality}. Specifically, $J_{2,1}$ and $J_{2,2}$ are the linear combinations of suitably chosen correlations defined as 
\begin{eqnarray}
\label{bl21}
\nonumber
J_{2,1}=\langle(A_{1}+A_{2}) B_{1} (C_{1}+C_{2})\rangle\\
J_{2,2}=\langle(A_{1}-A_{2}) B_{2} (C_{1}-C_{2})\rangle,
\end{eqnarray}
where $A_{x}$ denotes observables corresponding to the input $x$ of the  Alice $(x=1,2)$ and $C_{z}$ denotes observables corresponding to the input $z$ of the  Charlie $(z=1,2)$ and  $\langle{A_{x}B_{y}C_{z}}\rangle = \sum_{a,b,c}(-1)^{a+b+c}\hspace{1mm}P(a,b,c|x,y,z)$.

The assumption of independent sources is crucial to deriving the bilocality inequality in Eq. (\ref{Delbl22}). To satisfy the analogous condition in quantum theory, one may take the independent sources by assuming two entangled states that originated from the independent sources $S_{1}$ and $S_{2}$. It is shown \cite{Cyril2012} that the optimal quantum value $(\beta_{2})_{Q}^{opt}=2\sqrt{2} >(\beta_{2})_{bl}$. For this, Alice's (Charlie's) two observables are needed to be anticommuting and Bob's observables ${B}_{1}$ and ${B}_{2}$ to be commuting. Note here that the optimal quantum violation is commonly derived in literature by assuming two two-qubit systems. Here we provide a dimension independent derivation of the optimal quantum violation of n-locality inequalities without using the dimension of the system. For this, we follow an elegant SOS approach initiated in our recent work \cite{munshi2021}.

\section{Bilocality scenario in linear-chain network for three independent sources  $n=3$}
\begin{figure}[ht]
\centering
{{\includegraphics[width=1\linewidth]{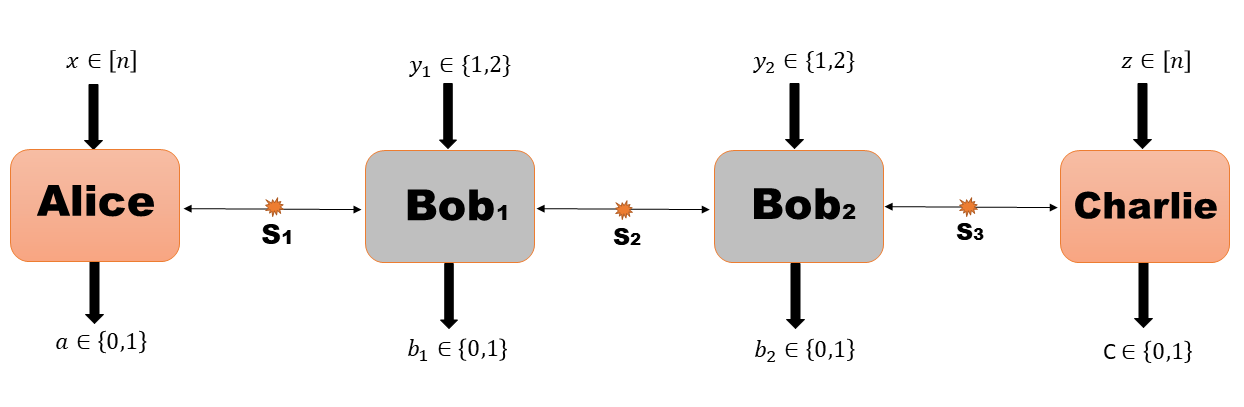}}}
\caption{Trilocal chain-shaped network scenario.}
\label{fig:my_labell}
\end{figure}
We first consider the trilocality scenario in linear-chain network. We propose a suitable trilocality inequality and derive its optimal quantum violation. As depicted in Fig. \ref*{fig:my_labell},  the trilocality scenario in linear chain-network configuration for $n=3$ features three independent sources $S_1$, $S_2$ and $S_3$ and  total four parties including two edge parties Alice and Charlie and two central parties $Bob_1$  and $Bob_2$. While source $S_1$ sends physical systems to Alice and $Bob_1$, the source $S_2$ sends particles to $Bob_1$ and $Bob_2$ and the source $S_3$ sends particles to $Bob_2$ and Charlie.  We assume Alice (Charlie) receives same number of inputs which is equal to the number of independent sources. Alice (Charlie) performs  binary outcome measurements according to the inputs $x\in\{1,2,3\}$ $(z\in\{1,2,3\})$ and produces output $a\in\{0,1\}$ $(c\in\{0,1\})$. Each Bob performs two dichotomic measurements upon receiving inputs  $y_1$, $y_2$ $\in\{1,2\}$ and produce output  $b_1$, $b_2\in\{0,1\}$ respectively. $Bob_1$ and $Bob_2$ performs joint measurement on the system they receive from sources $S_1$, $S_2$ and $S_3$ which are assumed to be independent. 

We assume three different hidden variables $\lambda_{1}$, $\lambda_{2}$ and $\lambda_{3}$ correspond to the sources $S_1$, $S_2$ and $S_3$ respectively having distribution $\rho (\lambda_{1},\lambda_{2},\lambda_{3})$ is assigned by the local model. Since $S_1$, $S_2$ and $S_3$ are assumed to be independent, the joint distribution $\rho (\lambda_{1},\lambda_{2},\lambda_{3})$ can be factorised as $\rho (\lambda_{1},\lambda_{2},\lambda_{3})$ = $\rho_{1}(\lambda_{1})$  $\rho_{2}(\lambda_{2})$  $\rho_{3}(\lambda_{3})$,   and $\rho_{1}{(\lambda_{1})}$, $\rho_{2}{(\lambda_{2})}$ and $\rho_{3}{(\lambda_{3})}$ are independent distribution that satisfy $\int d\lambda_{1}\rho_{1}{(\lambda_{1})}=1$, $\int d\lambda_{2}\rho_{2}{(\lambda_{2})}=1$ and $\int  d\lambda_{3}\rho_{3}{(\lambda_{3})}=1$. This constitute the trilocality condition in a network scenario  and the joint probability can then be written as
\begin{eqnarray}
	\nonumber
	P(a, b_1b_2, c|x,y_{1} y_{2},z)=\int\int d\lambda_{1} d\lambda_{2} d\lambda_{3}\hspace{2mm}\rho_{1}(\lambda_{1})\hspace{1mm}\rho_{2}(\lambda_{2})\hspace{1mm}\rho_{3}(\lambda_{3})\\
	\nonumber
	\times P(a|x,\lambda_{1})P(b_1|y_{1},\lambda_{1},\lambda_{2})P(b_2|y_{2},\lambda_{2},\lambda_{3}) P(c|z,\lambda_{3}).\hspace{9pt}
\end{eqnarray}
 The outcome of Alice (Charlie) solely depend on $\lambda_{1}$ (  $\lambda_{3}$),  the outcomes of $Bob_{1}$ and $Bob_{2}$ depend on both ($\lambda_{1}$ , $\lambda_{2}$) and ($\lambda_{2}$ , $\lambda_{3}$) respectively. We propose a nonlinear bilocal inequality for $n=3$ case as 
\begin{equation}
	\label{Delta3}
	(\beta_{3})_{bl} \leq \sum\limits_{i=1}^{4}|{J}_{3,i}|^{1/2} \leq 6, 
\end{equation}
where we define $J_{3,i} $ s (with $i=1,2,3,4$) by taking suitable linear combinations of correlations are given by
\begin{eqnarray}
	\label{Beta6}
	\nonumber
	&&J_{3,1}=\hspace{1mm}{\langle}(A_{1}+A_{2}+A_{3})B_{1}^1B_{1}^2(C_{1}+C_{2}+C_{3}){\rangle},\\
	\nonumber
	&& J_{3,2}=\hspace{1mm}{\langle}(A_{1}+A_{2}-A_{3})B_{1}^1B_{2}^2(C_{1}+C_{2}-C_{3}){\rangle},\\
	&&J_{3,3}=\hspace{1mm}{\langle}(A_{1}-A_{2}+A_{3})B_{2}^1B_{1}^2(C_{1}-C_{2}+C_{3}){\rangle},\\ 
	\nonumber
	&&J_{3,4}=\hspace{1mm}{\langle}(-A_{1}+A_{2}+A_{3})B_{2}^1B_{2}^2(-C_{1}+C_{2}+C_{3}){\rangle}.
\end{eqnarray}
Using bilocality assumption and by defining $\langle{A_{x}}\rangle_{\lambda_{1}} = \sum_{a}(-1)^{a}  P(a|x)$ and $\langle{C_{z}}\rangle_{\lambda_{3}} = \sum_{c}(-1)^{c}  P(c|z)$ where $x,z\in \{1,2,3\}$  along with considering $|\langle{B^1_{1}}\rangle_{\lambda_{1},\lambda_{2}}$, $|\langle{B^2_{1}}\rangle_{\lambda_{1},\lambda_{2}}|\leq{1},$ we can write
\begin{eqnarray}
	\label{Beta1}
	|J_{3,1}|&\leq&\int{d}\lambda_{1}\rho_{1}(\lambda_{1})|\langle{A_{1}}\rangle_{\lambda_{1}}+\langle{A_{2}}\rangle_{\lambda_{1}}+\langle{A_{3}}\rangle_{\lambda_{1}}|\\
	\nonumber
	&\times&
	\int{d}\lambda_{3}\rho_{3}(\lambda_{3})|\langle{C_{1}}\rangle_{\lambda_{3}}+\langle{C_{2}}\rangle_{\lambda_{3}}+\langle{C_{3}}\rangle_{\lambda_{3}}|. \hspace{5mm}
\end{eqnarray}
The term $|J_{3,2}|$, $|J_{3,3}|$ and $|J_{3,4}|$ given in  Eq. (\ref{Beta6}) can also be written in similar manner. Using the inequality
\begin{equation}
	\label{Beta7}
	\sum\limits_{i=1}^{2^{n-1}}\sqrt{r_{i}s_{i}}\leq\sqrt{\sum\limits_{i=1}^{2^{n-1}}r_{i}}\sqrt{\sum\limits_{i=1}^{2^{n-1}}s_{i}}
\end{equation} 
for $r_{i}, s_{i} \geq 0$, $i\in[4]$, we find from Eq.(\ref{Delta3}) that 
\begin{equation}
	(\beta_3)_{bl}\leq \sqrt{\int{d}\lambda_{1}\rho_{1}(\lambda_{1})\eta_A }\times \sqrt{\int{d}\lambda_{3}\rho_{3}(\lambda_{3})\eta_C },
\end{equation}
where \begin{eqnarray} \eta_A &=& |{\langle A_1\rangle}_{\lambda_{1}}+{\langle A_2\rangle}_{\lambda_{1}}+{\langle A_3\rangle}_{\lambda_{1}}|+|{\langle A_1\rangle}_{\lambda_{1}}+{\langle A_2\rangle}_{\lambda_{1}}-{\langle A_3\rangle}_{\lambda_{1}}|\\
\nonumber
&+&|{\langle A_1\rangle}_{\lambda_{1}}-{\langle A_2\rangle}_{\lambda_{1}}+{\langle A_3\rangle}_{\lambda_{1}}|+|-{\langle A_1\rangle}_{\lambda_{1}}+{\langle A_2\rangle}_{\lambda_{1}}+{\langle A_3\rangle}_{\lambda_{1}}|
\end{eqnarray}
\begin{eqnarray}
    \eta_C& =& |{\langle C_1\rangle}_{\lambda_{3}}+{\langle C_2\rangle}_{\lambda_{3}}+{\langle C_3\rangle}_{\lambda_{3}}|+|{\langle C_1\rangle}_{\lambda_{3}}+{\langle C_2\rangle}_{\lambda_{3}}-{\langle C_3\rangle}_{\lambda_{3}}|\\
    \nonumber&+&|{\langle C_1\rangle}_{\lambda_{3}}-{\langle C_2\rangle}_{\lambda_{3}}+{\langle C_3\rangle}_{\lambda_{3}}|+|-{\langle C_1\rangle}_{\lambda_{3}}+{\langle C_2\rangle}_{\lambda_{3}}+{\langle C_3\rangle}_{\lambda_{3}}|.
\end{eqnarray}

Since all the observables are dichotomic having values $\pm1$, it is simple to check that $\eta_A$=$\eta_C\leq 6$. Integrating over $\lambda_{1}$ and $\lambda_{3}$ we obtain $(\beta_3)_{bl}\leq 6$, as claimed in Eq. (\ref{Delta3}).

As already mentioned, to derive the optimal value of $(\beta_{3})_Q,$ we use an elegant SOS approach. For this we  define  a positive semi-definite operator $\gamma_3$, that can be expressed as \
	\begin{equation}
{\left\langle\gamma_3\right\rangle}_Q  = -(\beta_{3})_Q + \tau_3.	\end{equation}
Here $\tau_3$ is the optimal value, can be obtained when ${\left\langle\gamma_3\right\rangle}_Q $ is equal to zero. This can be proven by considering a set of suitable positive operators $M_{3,i}$ which is a polynomial function of $A_x$, $B_y$, and $C_z$ so that
	\begin{equation}
	\label{apx1}
		\langle\gamma_{3}\rangle=\sum\limits_{i=1}^{4}\frac{{(\omega_{3,i}})^{\frac{1}{2}}}{2 }\langle\psi|(M_{3,i})^{\dagger}M_{3,i}|\psi\rangle,
	\end{equation}
	where $\omega_{3,i}$ is a suitable positive numbers and  $\omega_{3,i}$ = $\omega^A_{3,i}  \omega^C_{3,i}$. We choose a suitable set of positive operators $M_{3,i}$ as
	\begin{widetext}
	\begin{eqnarray}
	\label{apx2}
		M_{3,1}{|\psi\rangle}_{AB_{1}B_{2}C} = \sqrt{\bigg|\bigg(\frac{A_1 + A_2 + A_3}{\omega^A_{3,1}}\otimes \frac{C_1 + C_2 + C_3}{\omega^C_{3,1}}\bigg){|\psi\rangle}_{AB_{1}B_{2}C}\bigg|} - \sqrt{|B^1_1\otimes B^2_1{|\psi\rangle}_{AB_{1}B_{2}C}|}\\
		\label{apx3}
		M_{3,2}{|\psi\rangle}_{AB_{1}B_{2}C} = \sqrt{\bigg|\bigg(\frac{A_1 + A_2 - A_3}{\omega^A_{3,2}}\otimes \frac{C_1 + C_2 - C_3}{\omega^C_{3,2}}\bigg){|\psi\rangle}_{AB_{1}B_{2}C}\bigg|} - \sqrt{|B^1_1\otimes B^2_2{|\psi\rangle}_{AB_{1}B_{2}C}|}
      \end{eqnarray}
  \begin{eqnarray}
		\label{apx4}
		M_{3,3}{|\psi\rangle}_{AB_{1}B_{2}C} = \sqrt{\bigg|\bigg(\frac{A_1 - A_2 + A_3}{\omega^A_{3,3}}\otimes \frac{C_1 - C_2 + C_3}{\omega^C_{3,3}}\bigg){|\psi\rangle}_{AB_{1}B_{2}C}\bigg|} - \sqrt{|B^1_2\otimes B^2_1{|\psi\rangle}_{AB_{1}B_{2}C}|},\\
		\label{apx5}
		M_{3,4}{|\psi\rangle}_{AB_{1}B_{2}C} = \sqrt{\bigg|\bigg(\frac{-A_1 + A_2 + A_3}{\omega^A_{3,4}}\otimes \frac{-C_1 + C_2 + C_3}{\omega^C_{3,4}}\bigg){|\psi\rangle}_{AB_{1}B_{2}C}\bigg|} - \sqrt{|B^1_2\otimes B^2_2{|\psi\rangle}_{AB_{1}B_{2}C}|},
	\end{eqnarray}
		\end{widetext}
	where $\omega^A_{3,1}$ = ${|| (A_1 + A_2 + A_3){|\psi\rangle}_{AB_{1}B_{2}C}||}_2$ , $\omega^C_{3,1}$= ${|| (C_1 + C_2 + C_3){|\psi\rangle}_{AB_{1}B_{2}C}||}_2$ and similarly for other $\omega^A_{3,i}$ , $\omega^C_{3,i}$, $i = 1,2,3,4$. Here ${||.||}_2$ denotes the norm of the vector. Plugging Eqs. (\ref  {apx2}-\ref {apx5})  in Eq. (\ref{apx1}), we get ${\left\langle\gamma_3\right\rangle}_Q $ = $-(\beta_{3})_Q$ + $\sum_{i=1}^{4}\sqrt{\omega_{3,i}}$. The optimal quantum value is obtained if ${\left\langle\gamma_3\right\rangle}_Q $ = 0, implying that $\forall i$, $M_{3,i}{|\psi\rangle}_{AB_{1}B_{2}C}$ = 0. Hence,
	\begin{equation}
		{(\beta_{3})}^{opt}_Q = max\hspace{1mm} \bigg(\sum_{i=1}^{4}\sqrt{\omega_{3,i}}\bigg).
	\end{equation} 
	
	As defined,  $\omega^A_{3,1}$ = ${|| (A_1 + A_2 + A_3){|\psi\rangle}_{AB_{1}B_{2}C}||}_2$ = $\sqrt{3 + \langle (\{A_1,A_2\} + \{A_2,A_3\} + \{A_3,A_1\})\rangle}$,  and similarly for other $\omega^A_{3,i}$ , $\omega^C_{3,i}$, $i = 1,2,3,4$.
	Since $\omega_{3,i}$ = $\omega^A_{3,i}$ . $\omega^C_{3,i}$, by using the inequality $\sum\limits_{i=1}^{4}\sqrt{r_{i}s_{i}}\leq\sqrt{\sum\limits_{i=1}^{4}r_{i}}\sqrt{\sum\limits_{i=1}^{4}s_{i}}$ (for $r_{i}, s_{i} \geq 0$,$i=1,2,3,4$), we get $\sum_{i=1}^{4} \sqrt{\omega_{3,i}}\leq\sqrt{\sum_{i=1}^{4} \omega^A_{3,i}}$
	$\sqrt{\sum_{i=1}^{4} \omega^C_{3,i}}$ . Finally, by using the identity $\sqrt{b+a} + \sqrt{b-a} = \sqrt{2b + 2\sqrt{b^2 - a^2}}$, we obtain
		\begin{widetext}
	\begin{eqnarray}
		\nonumber
		\sum_{i=1}^{4} \sqrt{\omega_{3,i}}\leq\bigg[\sqrt{2\bigg(3+\langle\{A_1,A_2\}\rangle\bigg) + 2\sqrt{\bigg(3+\langle\{A_1,A_2\}\rangle\bigg)^2-\bigg(\langle\{A_2,A_3\}\rangle + \langle\{A_3,A_1\}\rangle\bigg)^2}}\\
		+\sqrt{2\bigg(3+\langle\{A_1,A_2\}\rangle\bigg) + 2\sqrt{\bigg(3-\langle\{A_1,A_2\}\rangle\bigg)^2-\bigg(\langle\{A_2,A_3\}\rangle - \langle\{A_3,A_1\}\rangle\bigg)^2}}\hspace{1mm}\bigg]^{\frac{1}{2}}\\
		\nonumber
		\bigg[\sqrt{2\bigg(3+\langle\{C_1,C_2\}\rangle\bigg) + 2\sqrt{\bigg(3+\langle\{C_1,C_2\}\rangle\bigg)^2-\bigg(\langle\{C_2,C_3\}\rangle + \langle\{C_3,C_1\}\rangle\bigg)^2}}\\
		\nonumber
		+\sqrt{2\bigg(3-\langle\{C_1,C_2\}\rangle\bigg) + 2\sqrt{\bigg(3-\langle\{C_1,C_2\}\rangle\bigg)^2-\bigg(\langle\{C_2,C_3\}\rangle - \langle\{C_3,C_1\}\rangle\bigg)^2}}\hspace{1mm}\bigg]^{\frac{1}{2}}.
	\end{eqnarray}
	\end{widetext}
	Clearly, ${(\beta_{3})}^{opt}_Q = max\hspace{1mm} \bigg(\sum_{1}^{4}\sqrt{\omega_{3,i}}\bigg)$ = $4\sqrt 3$ when $A_1$, $A_2$ and $A_3$ are mutually anticommuting and $C_1$, $C_2$ and $C_3$ are also mutually anticommuting. 
 
 It is straightforward to find the Bob's observables from $ M_{3,i}{|\psi\rangle}_{AB_{1}B_{2}C}$ = $0$, $\forall i$. This in turn, fixes the state ${|\psi\rangle}_{AB_{1}B_{2}C}= {|\psi\rangle}_{AB_{1}}\otimes{|\psi\rangle}_{B_{1}B_{2}}\otimes {|\psi\rangle}_{B_{2}C}$ with ${|\psi\rangle}_{ A B_{1}}$, ${|\psi\rangle}_{B_{1}B_{2}}$ and ${|\psi\rangle}_{B_{2}C}$ to be two-qubit maximally entangled states. The same approach will be used  to derive the optimal quantum violation of the family of $n$-local inequalities. However, for arbitrary $n>3$ scenario, two-qubit entangled states are not enough to obtain optimal quantum violation of $n$-locality inequalities and one has to invoke higher-dimensional  entangled state.

\section{$n$-Locality Generalisation in chain network}

\begin{figure*}[ht]
	\centering
	{{\includegraphics[width=0.9\linewidth]{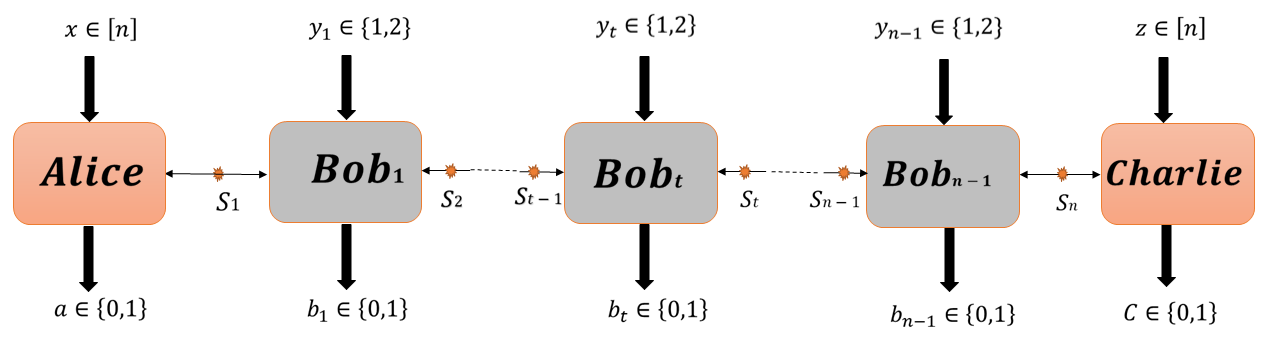}}}
	\caption{The n-local chain-shaped network scenario is depicted. There are  $n$ independent sources and $n+1$ parties that includes two edge parties (Alice and Charlie) and $n-1$ central parties $Bob_m$ ($m$=1,2,.....,$n-1$).}
	\label{fig:my_label}
\end{figure*}
We now generalize the $n$-local linear-chain network scenario depicted in 
Fig. \ref*{fig:my_label}. There are $n$ independent sources where each source distributes physical subsystems to adjacent parties. For example, source $S_1$ distributes subsystems to Alice and $Bob_1$ and so on. There are $n+1$ parties including two edge parties Alice and Charlie, and $n-1$ central parties  $Bob_m$ $(m = 1,2,.......,n-1)$ organized in a linear-chain network. The edge party Alice (Charlie) receives inputs $x\in\{ 1,2,.....,n\}$($z\in\{ 1,2,.....,n\}$), and produces output $a\in\{0,1\}$($c\in\{0,1\}$). Except the edge parties, each of the $n-1$ parties receives only two inputs $y_1$,$y_2$,.......,$y_{n-1}\in\{1,2\}$, produces respective outputs $b_1$,$b_2$,......,$b_{n-1}\in\{0,1\}$. Then, for each central party there will be a joint measurement on two subsystems simultaneously. Let us propose a generalized $n$-locality expression as
\begin{equation}
	\label{Beta8}
	\beta_n = \sum_{i=1}^{2^{n-1}} \sqrt{|J_{n,i}|},
\end{equation}
where $J_{n,i}$ is the linear combination of suitable correlations are defines as
\begin{equation}
	\label{Beta5}
	J_{n,i}=\sum_{x=1}^{n}(-1)^{y_{x}^{i}}\langle{A_{x}}\rangle\otimes\langle B^{1}_{y^k_1}\otimes .....\otimes B^{n-1}_{y^k_{n-1}}\rangle\otimes\sum_{z=1}^{n}(-1)^{y_{z}^{i}}\langle{C_{z}}\rangle.
\end{equation}
Here ${y_{x}^{i}}$ or ${y^i_z}$ takes value either 0 or 1 which are fixed through the encoding scheme used in random-access-code \cite{Ghorai2018,AKP2020,Asmita}. This will fix $1$ or $-1$ values of $(-1)^{y_{x}^{i}}$ and $(-1)^{y_{z}^{i}}$ in  Eq. (\ref{Beta5}) for a given $i$. Let us consider a random variable $y^{\delta}\in \{0,1\}^{n}$ with $\delta\in \{1,2...2^{n}\}$. Each element of the bit string can be written $y^{\delta}=y^{\delta}_{x_{1}=1} y^{\delta}_{x_{1}=2} y^{\delta}_{x_{1}=3} .... y^{\delta}_{x_{1}=n}$, for  example, if $y^{\delta} = 011...00$ then $y^{{\delta}}_{x_{1}=1} =0$, $y^{{\delta}}_{x_{1}=2} =1$, $y^{{\delta}}_{x_{1}=3} =1$ and so on. 

For our purpose here, we denote $2^{n-1}$ numbers of length $n$ binary string as $y^{i}$ that have $0$ in its first bit in $y^{\delta}$ with $i\in \{1,2...2^{n-1}\}$. As an example, for $n=3$, we have $y^{\delta}\in \{000,001,010,100,011,101,110,111\}$ with $\delta =1,2,...,8$. We then denote $y^{i} \in \{000,001,010,011\}$ with $y^{1}=000$, $y^{2}=001$, $y^{3}=010$ and  $y^{4}=011$. This also means  $y^{1}_{x_{1}=1}=0,  y^{1}_{x_{1}=2}=0,   y^{1}_{x_{3}=1}=0$ and so on.

Now, for $y^k\in\{0,1\}^{n-1}$  with $k\in \{1,2...2^{n-1}\}$, taking above example of $n=3$ we have $y^k\in\{00,01,10,11\}$ with $k\in\{1,2,3,4\}$. We can then say   $y^{1}_1=0$ and $y^{1}_2=0$, $y^{2}_1=0$ and $y^{2}_2=1$. Putting $n=2$ and $n=3$ in  Eq. (\ref{Beta5}), we can recover the expressions present  in Eqs. (\ref{bl21}) and (\ref{Beta6}) respectively.

 We consider that for each $Bob_m$ with  $m\in\{1,2,....,n-1\}$, receives input $y\in\{1,2\}$ and the output  depends on the physical states $\lambda_{t}$ and $\lambda_{t+1}$, Whereas  Alice (Charlie) receives input  $x\in[n]$ ($z\in[n]$), and the outputs  depends on  $\lambda_{1}$ ($\lambda_{n}$) . Taking into consideration of $n$ independent sources, we can factorize the probability distribution as $\rho(\lambda_{1},.....\lambda_{n}) = \prod_{t=1}^{n} \rho_{t}(\lambda_{t})$. We can write $J_{n,i}$ in a $n$-level model as 
\begin{eqnarray}
	\nonumber
	|J_{n,i}|=\int\int && d\lambda_{1}\hspace{1mm}\rho_{1}(\lambda_{1}),..., d\lambda_{n}\hspace{1mm}\rho_{n}(\lambda_{n})
	\times\bigg(\sum_{x=1}^{n}(-1)^{y_{x}^{i}}\langle{A_{x}}\rangle_{\lambda_{1}}\bigg)\\
	\nonumber
	&&\bigg(\langle{B_{1}}\rangle_{\lambda_{1}\lambda_{2}},...,\langle{B_{n-1}}\rangle_{\lambda_{n-1}\lambda_{n}}\bigg) \bigg(\sum_{z=1}^{n}(-1)^{y_{z}^{i}}\langle{C_{z}}\rangle_{\lambda_{n}}\bigg).\\
\end{eqnarray}
Using the fact that\hspace{1mm} $|\langle{B_{1}}\rangle_{\lambda_{1}\lambda_{2}},....,\langle{B_{n-1}}\rangle_{\lambda_{n-1}\lambda_{n}}|\leq {1}$, we write 
\begin{equation}
	\label{Beta13}
	|J_{n,i}|\leq\int\int d\lambda_{1}\hspace{1mm}\rho_{1}(\lambda_{1}),...,d\lambda_{n}\hspace{1mm}\rho_{n}(\lambda_{n})\times\bigg|J^{A}_{n,i}\cdot J^{C}_{n,i}\bigg|.
\end{equation}
To avoid notational clumsiness, we denote
\begin{eqnarray}
|J^{A}_{n,i}| = \bigg(\sum_{x=1}^{n}(-1)^{y_{x}^{i}}\langle{A_{x}}\rangle_{\lambda_{1}}\bigg); \ \ |J^{C}_{n,i}| = \bigg(\sum_{x=1}^{n}(-1)^{y_{z}^{i}}\langle{C_{z}\rangle_{\lambda_{n}}\bigg)}.
\end{eqnarray}
Using the condition $\int d\lambda_{2}\rho_{2}{(\lambda_{2})}=1$, $\int d\lambda_{3}\rho_{3}{(\lambda_{3})}=1$,...,$\int d\lambda_{n-1}\rho_{n-1}{(\lambda_{n-1})}=1$ we can further simplify Eq. (\ref{Beta13}) , we get 
\begin{equation}
	\sqrt{|J_{n,i}|}\leq\sqrt{\int d\lambda_{1}\rho_{1}(\lambda_{1})\hspace{1mm}|J^{A}_{n,i}| \times\int d\lambda_{n}\rho_{n}(\lambda_{n})\hspace{1mm}|J^{C}_{n,i}|}.
\end{equation}
Further using the  inequality Eq. (\ref{Beta7}), the expression $\beta_{n}$ in Eq. (\ref{Beta8}) becomes :
\begin{equation}
\label{Beta12}
(\beta_{n})\leq\sqrt{\int d\lambda_{1}\rho_{1}(\lambda_{1})\sum_{i=1}^{2^{n-1}}|J^{A}_{n,i}|}\times \sqrt{\int d\lambda_{n}\rho_{n}(\lambda_{n})\sum_{i=1}^{2^{n-1}}|J^{C}_{n,i}|}.
\end{equation}
For notational convenience, we further denote
\begin{eqnarray}
	\label{Beta9}
	\sum_{i=1}^{2^{n-1}}|J^{A}_{n,i}| = \sum_{i=1}^{2^{n-1}}\bigg|\bigg(\sum_{x=1}^{n}(-1)^{y_{x}^{i}}\langle{A_{x}}\rangle_{\lambda_{1}}\bigg)\bigg| = \alpha_A,\\
	\label{Beta10}
	\sum_{i=1}^{2^{n-1}}|J^{C}_{n,i}| = \sum_{i=1}^{2^{n-1}}\bigg|\bigg(\sum_{z=1}^{n}(-1)^{y_{z}^{i}}\langle{C_{z}\rangle_{\lambda_{n}}\bigg)}\bigg| = \alpha_C.
\end{eqnarray}
Both Eqs. (\ref{Beta9} - \ref{Beta10}) admits the same value and we can write $\alpha_A$ = $\alpha_C$ = $\alpha$ as we have taken the same encoding scheme for both Alice and Charlie. Putting $\alpha_A$ and $\alpha_C$ in Eq. (\ref{Beta12}) and integrating, we get that the quantity $\alpha$ is the upper bound in a $n$-local model whose maximum value has to be determined for arbitrary $n$.  We obtain classical bound for $n$-locality inequality is given by
\begin{equation}
(\beta_{n})_{nl}\leq\alpha = \sum\limits_{l=0}^{\lfloor\frac{n}{2}\rfloor}\binom{n}{l}(n-2l).
\end{equation}
Here subscript $nl$ denotes $n$-local scenario. This was derived by us in \cite{munshi2021} for a different context for star-network configuration. We can recover the respective bilocal inequality for $n=2$, i.e., $(\beta_{2})\leq 2$ and trilocal inequality for $n=3$, i.e., $(\beta_{3})\leq 6$ respectively.
\subsection{Quantum-violation of $n$ locality inequality}
To find the optimal quantum value of the expression $\beta_{n}$ in Eq. (\ref{Beta12}), we use an improved version of SOS approach, so that $(\beta_{n})_Q\leq \tau_{n}$ for all possible quantum states. This is equivalent to showing that there is a positive semi-definite operator $(\gamma_n)_Q\geq 0$, which can be expressed as ${\langle\gamma_n\rangle}_Q$ $=$ $-(\beta_n)_Q + \tau_{n}$. This can be proven by considering a set of suitable positive operators $M_{n,i}$ which are polynomial functions of $A_x$, $C_z$ and ($B^{1}_{y^{k}_{1}}$$B^{2}_{y^{k}_{2}}$....$B^{n-1}_{y^{k}_{n-1}}$) such that
\begin{equation}
	\label{gammanm}
	\langle\gamma_{n}\rangle=\sum\limits_{i=1}^{2^{n-1}}  \frac{{(\omega_{n,i}})^{\frac{1}{2}}}{2 }\langle\psi|(M_{n,i})^{\dagger}M_{n,i}|\psi\rangle,
\end{equation}
where  $\omega_{n,i}$ is a positive number with $\omega_{n,i}=(\omega^A_{n,i})  (\omega^C_{n,i})$. The optimal quantum value of $(\beta_n)_Q$ is obtained if $(\gamma_n)_Q = 0$, i.e.,
\begin{equation} 
	\forall {i}, \hspace{3mm}     M_{n,i}\ket{\psi}_{AB_1B_2....B_{n-1}C} = 0,
\end{equation}
Where $\ket{\psi}_{AB_1B_2....B_{n-1}C} = \ket{\psi}_{AB_1}\otimes\ket{\psi}_{B_1B_2}\otimes...\otimes\ket{\psi}_{B_{n-1}C}$ and $\ket{\psi}_{AB_1}$, $\ket{\psi}_{B_1B_2}$,..., and $\ket{\psi}_{B_{n-1}C}$ are originating from independent sources $S_{1}$,  $S_{2}$,.., $S_{n}$ respectively. For notational convenience we denote $\ket{\psi}_{AB_1B_2....B_{n-1}C} = \ket{\psi}$. The operators $M_{n,i}\ket{\psi}$ are suitably chosen as 

\begin{equation}
	\label{Mnmi}
	M_{n,i}\ket{\psi}=\frac{1}{{(\omega_{n,i})}^\frac{1}{2}}\bigg|\bigg[\sum\limits_{x,z=1}^{n}(-1)^{y^{i}_{x}} A_{x}\otimes\hspace{1mm} (-1)^{y^{i}_{z}} C_{z}\hspace{1pt}\bigg]\ket{\psi}\bigg|^{\frac{1}{2}}-|B_{i}\ket{\psi}|^{\frac{1}{2}},
\end{equation}

where
\begin{eqnarray}
	(\omega^{A}_{n,i})=\bigg|\bigg|\bigg[\sum\limits_{x=1}^{n}(-1)^{y^{i}_{x}} A_{x}\bigg]\ket{\psi}\bigg|\bigg|_{2},\\
	(\omega^{C}_{n,i})=\bigg|\bigg|\bigg[\sum\limits_{z=1}^{n}(-1)^{y^{i}_{z}} C_{z}\bigg]\ket{\psi}\bigg|\bigg|_{2}.
\end{eqnarray}
Also here $B_i$ where ($i$=1,2,....,$2^{n-1}$) signifies sequential combination of all the central observer for different value for $i$.  For $i$ = 1, the combination works as $ B_1 = B^1_1\otimes B^2_1\otimes....\otimes B^{n-2}_1\otimes B^{n-1}_1.$  For $i$ = 2, the combination works as $ B_2 = B^1_1\otimes B^2_1\otimes....\otimes B^{n-2}_1\otimes B^{n-1}_2. $ For $i$ = $2^{n-1}$,the combination works as $ B_{2^{n-1}} = B^1_2\otimes B^2_2\otimes....\otimes B^{n-2}_2\otimes B^{n-1}_2.$\\ 
For notational convenience let $\bigg[\sum\limits_{x=1}^{n}(-1)^{y^{i}_{x}} A_{x}\hspace{1pt}\bigg]=Y_{i}^{A}$ and  $\bigg[\sum\limits_{z=1}^{n}(-1)^{y^{i}_{z}} C_{z}\hspace{1pt}\bigg]=Y_{i}^{C}$. Also note that, $(Y_{i}^{A})^{\dagger}=Y_{i}^{A}$ and $(Y_{i}^{C})^{\dagger}=Y_{i}^{C}$. We can also write 

\begin{eqnarray}
	\label{ymi} 
	\left((\omega^A_{n,i})\right)^{2}=\langle\psi| (Y_{i}^{A})^{\dagger}Y_{i}^{A}|\psi\rangle= \langle\psi| (Y_{i}^{A})^{2}|\psi\rangle,\\
	\label{ymii}
	\left((\omega^C_{n,i})\right)^{2}=\langle\psi| (Y_{i}^{C})^{\dagger}Y_{1}^{C}|\psi\rangle= \langle\psi| (Y_{i}^{C})^{2}|\psi\rangle. 
\end{eqnarray}

Eq. (\ref{Mnmi}) can then be written as 
\begin{eqnarray}
	M_{n,i}|\psi\rangle=\frac{1}{(\omega_{n,i})^{\frac{1}{2}}}\bigg|\bigg[Y_{i}^{A}\otimes Y_{i}^{C}|\psi\rangle\bigg]\bigg|^{\frac{1}{2}}-|B_{i}|\psi\rangle|^{\frac{1}{2}},
\end{eqnarray}
where $i\hspace{1mm} \forall\hspace{2mm} \{1,2,.....,2^{n-1}\}$.
This provides
\begin{eqnarray}
	\nonumber	
	&&\langle\psi|(M_{n,i})^{\dagger}M_{n,i}|\psi\rangle=\frac{1}{(\omega_{n,i})}\left(\langle\psi|\left(Y_{i}^{A}\otimes Y_{i}^{C} \right)^{2}|\psi\rangle\right)\\
	&&-\frac{2}{(\omega_{n,i})^{\frac{1}{2}}}\bigg|\langle\psi|Y_{i}^{A}\otimes Y_{i}^{C}\otimes  B_{i}|\psi\rangle\bigg|^{\frac{1}{2}}+\left(\langle\psi|(B_{i})^{2}|\psi\rangle\right)\hspace{9pt}\\
	&&=\label{kkm,i}
	2-2\bigg|\frac{1}{(\omega_{n,i})}\langle\psi|Y_{i}^{A}\otimes Y_{i}^{C}\otimes B_{i}|\psi\rangle\bigg|^{\frac{1}{2}}.
\end{eqnarray}
where we have used Eqs. (\ref{ymi}-\ref{ymii}) and $(B_{i})^{2}=\mathbb{I}$. Putting Eq. (\ref{kkm,i}) in Eq. (\ref{gammanm}), we get

\begin{eqnarray}
	\langle\gamma_{n}\rangle_{Q}&&=\sum\limits_{i=1}^{2^{n-1}}(\omega_{n,i})^{\frac{1}{2}}-\sum\limits_{i=1}^{2^{n-1}}\left[\bigg|\langle\psi|Y_{i}^{A}\otimes Y_{i}^{C}\otimes B_{i}|\psi\rangle\bigg|^{\frac{1}{2}}\right]\\
	&& =\sum\limits_{i=1}^{2^{n-1}}(\omega_{n,i})^{\frac{1}{2}}-(\beta_{n})_{Q}.
\end{eqnarray}
Since ${\langle\gamma_{n}\rangle}_Q\geq 0$. we have found the value of $(\beta_n)_Q$ as

\begin{align}
	\label{optbnm}
	(\beta_{n})_{Q}^{opt} =max\left( \sum\limits_{i=1}^{2^{n-1}}(\omega_{n,i})^{\frac{1}{2}}\right)
\end{align} 
and the explicit condition can be found from the fact that 	$\forall i, \ \ \ M_{n,i}|\psi\rangle_{A_{1}A_{2}\cdots A_{n}B}=0$. Using Eq. (\ref{Beta7}),
we can write $\sum\limits_{i=1}^{2^{n-1}}(\omega^{A}_{n,i}\cdot\omega^{C}_{n,i})^{\frac{1}{2}}\leq\bigg(\sum\limits_{i=1}^{2^{n-1}}(\omega^{A}_{n,i})\bigg)^{\frac{1}{2}}\cdot                  \hspace{2mm}\bigg(\sum\limits_{i=1}^{2^{n-1}}(\omega^{C}_{n,i})\bigg)^{\frac{1}{2}}$. Since $A_{x}$ and $C_{x}$ are dichotomic, the quantity $(\omega^{A}_{n,i})$ and $(\omega^{C}_{n,i})$ can be written as,    
\begin{widetext}
\begin{eqnarray}
\label{omega}
(\omega^{A}_{n,i})&=&||\sum\limits_{x=1}^{n} (-1)^{y^i_{x}} A_{x}\ket{\psi}||_{2}=\bigg(\langle\psi|\bigg[\sum\limits_{x=1}^{n} (-1)^{y^i_{x}} A_{x}\bigg]^{\dagger}\bigg[\sum\limits_{x=1}^{n} (-1)^{y^i_{x}}A_{x}\bigg]|\psi\rangle\bigg)^{1/2}\\
\nonumber
&&=\Big[n+\Big\langle\{(-1)^{y^{i}_{1}}A_{1},
\sum\limits_{x=2}^{n}(-1)^{y^{i}_{x}} A_{x}\}]\Big\rangle+\Big\langle\{(-1)^{y^i_2} A_{2}, \sum\limits_{x=3}^{n}(-1)^{y^{i}_{x}} A_{x}\}\Big\rangle +\cdots +\Big\langle\{(-1)^{y^i_{n-1}} A_{(n-1)}, (-1)^{y^i_n} A_{n}\}\Big)\Big\rangle\Big]^{1/2}
\end{eqnarray}
\end{widetext}
and a similar expression can be written for $	(\omega^{C}_{n,i})$. Following the similar process used for $n=2$ and $n=3$, we derive $(\omega^{A}_{n,i})\leq \sqrt{n}$ and $(\omega^{C}_{n,i})\leq \sqrt{n}$ for every $i$ , and the equality holds only when each anticommutators is zero. 
We get from Eqs. (\ref{omega}) we have $\bigg(\sum\limits_{i=1}^{2^{n-1}}(\omega_{n,i})\bigg)\leq 2^{n-1}\sqrt{n}$ . Thus   Eq. (\ref{optbnm}) implies $(\beta_{n})_{Q} \leq 2^{n-1}\sqrt{n}$. Hence, the optimal quantum value  $(\beta_{n})_{Q}^{opt}$ can only be achieved when each edge parties, Alice and Charlie consider  $n$ number of anticommuting observables. Finally, from Eq. (\ref{optbnm}), the optimal quantum value of $(\beta_{n})_{Q}$ is 
\begin{align}
	\label{opt}
	(\beta_{n})_{Q}^{opt}= 2^{n-1}\sqrt{n}.
\end{align}
Clearly, $	(\beta_{n})_{Q}^{opt}> (\beta_{n})_{nl}$ for any value of $n$, thereby violating the $n$-locality inequality for linear-chain network. Each edge party requires an arbitrary $n$ number of mutually anticommuting observables thereby requiring higher-dimensional system when $n > 3$ to derive optimal quantum value. We derive that Alice and Bob$_{1}$ (also Charlie and Bob$_{n-1}$) have to share $\lfloor{\frac{n}{2}}\rfloor$ number of two-qubit entangled state to obtain the optimal quantum violation of $n$-locality inequality. 

\section{SUMMARY AND DISCUSSION}
We provided a nontrivial generalization of the $n$-locality scenario in the linear-chain network configuration. Such a network involves two edge parties (Alice and charlie), $n-1$ number of central parties $(Bob_1, Bob_2,...,Bob_{n-1})$ and $n$ number of independent sources. Instead of two dichotomic measurements per party as is commonly assumed, we considered that both edge parties perform $n$ number of binary-outcome measurements (equals to a number of independent sources), and each central party performs two measurements. We derived a generalized $n$-locality inequality and its optimal quantum violation in the scenario mentioned above.

We first provided a detailed analysis of the $n$-locality inequality for $n=3$ case and derived the trilocality inequality when the edge parties perform three binary-outcome measurements, and two central parties perform two dichotomic measurements. To obtain the optimal quantum value, we showed that Alice and Charlie have to share a two-qubit maximally entangled state with $Bob_1$ and $Bob_2$. The optimal quantum value is attained when observables of both Alice and Charlie are mutually anticommuting. Bob's observables and the required entangled states are also fixed by the optimization condition. We derived the optimal quantum violation using an elegant SOS approach without assuming the dimension of the system. 

We generalized our treatment for an arbitrary $n$ input scenario and derived the  $n$-locality inequality in linear-chain network configuration. Using the SOS approach again, we derived the optimal quantum violation of  $n$-locality inequality. We found that a two-qubit entangled state is not enough to obtain optimal quantum violation for $n>3$ as Alice and Charlie both require $n$ number of mutually anticommuting observables. We demonstrated that  Alice and Bob$_{1}$ as well as Charlie have to share $\lfloor{\frac{n}{2}}\rfloor$ number of two-qubit entangled state to obtain the optimal quantum violation of $n$-locality inequality.

\section* {Acnowledgements}
R.K. acknowledges UGC-CSIR NET-JRF
(Fellowship No. 16-6(Dec.2017)/2018(NET/CSIR)] for financial support. A.K.P. acknowledges the support from the project DST/ICPS/QuST/Theme 1/2019/4.


\begin{thebibliography}{0}%
\makeatletter
\providecommand \@ifxundefined [1]{%
 \@ifx{#1\undefined}
}%
\providecommand \@ifnum [1]{%
 \ifnum #1\expandafter \@firstoftwo
 \else \expandafter \@secondoftwo
 \fi
}%
\providecommand \@ifx [1]{%
 \ifx #1\expandafter \@firstoftwo
 \else \expandafter \@secondoftwo
 \fi
}%
\providecommand \natexlab [1]{#1}%
\providecommand \enquote  [1]{``#1''}%
\providecommand \bibnamefont  [1]{#1}%
\providecommand \bibfnamefont [1]{#1}%
\providecommand \citenamefont [1]{#1}%
\providecommand \href@noop [0]{\@secondoftwo}%
\providecommand \href [0]{\begingroup \@sanitize@url \@href}%
\providecommand \@href[1]{\@@startlink{#1}\@@href}%
\providecommand \@@href[1]{\endgroup#1\@@endlink}%
\providecommand \@sanitize@url [0]{\catcode `\\12\catcode `\$12\catcode
  `\&12\catcode `\#12\catcode `\^12\catcode `\_12\catcode `\%12\relax}%
\providecommand \@@startlink[1]{}%
\providecommand \@@endlink[0]{}%
\providecommand \url  [0]{\begingroup\@sanitize@url \@url }%
\providecommand \@url [1]{\endgroup\@href {#1}{\urlprefix }}%
\providecommand \urlprefix  [0]{URL }%
\providecommand \Eprint [0]{\href }%
\providecommand \doibase [0]{https://doi.org/}%
\providecommand \selectlanguage [0]{\@gobble}%
\providecommand \bibinfo  [0]{\@secondoftwo}%
\providecommand \bibfield  [0]{\@secondoftwo}%
\providecommand \translation [1]{[#1]}%
\providecommand \BibitemOpen [0]{}%
\providecommand \bibitemStop [0]{}%
\providecommand \bibitemNoStop [0]{.\EOS\space}%
\providecommand \EOS [0]{\spacefactor3000\relax}%
\providecommand \BibitemShut  [1]{\csname bibitem#1\endcsname}%
\let\auto@bib@innerbib\@empty
\end{thebibliography}%


\begin{thebibliography}{99}
	\bibitem{bell}J. Bell, \href{https://doi.org/10.1103/PhysicsPhysiqueFizika.1.195}{Physics, {\bf 1}, 195 (1964)}.
	\bibitem{brunnerreview} N. Brunner, D. Cavalcanti, S. Pironio, V. Scarani, and S. Wehner,  \href{https://doi.org/10.1103/RevModPhys.86.419}{Rev. Mod. Phys. {\bf 86}, 419 (2014)}.
	
	\bibitem{Cyril2012} C. Branciard, D. Rosset, N. Gisin, and S. Pironio, \href{https://journals.aps.org/pra/pdf/10.1103/PhysRevA.85.032119} {Phys. Rev. A {\bf 85}, 032119 (2012)}.
	\bibitem{Bran2010}  C. Branciard, N. Gisin, and S. Pironio,  \href{https://doi.org/10.1103/PhysRevLett.104.170401}{Phys. Rev. Lett. {\bf 104}, 170401 (2010)}.
	\bibitem{Gisin2017} N. Gisin, Q. Mei, A. Tavakoli, M. O. Renou, and N. Brunner, \href{https://doi.org/10.1103/PhysRevA.96.020304} {Phys. Rev. A {\bf 96}, 020304 (2017)}
	\bibitem{Frit2012}  T. Fritz, \href{https://iopscience.iop.org/article/10.1088/1367-2630/14/10/103001} {New J.
		Phys. {\bf 14}, 103001 (2012)}.
		\bibitem{Armi2014} A. Tavakoli, P. Skrzypczyk, D. Cavalcanti, and A. Acin,  \href{https://doi.org/10.1103/PhysRevA.90.062109}  {Phys. Rev. A {\bf 90}, 062109 (2014)}.
	\bibitem{Tavakoli2016}A. Tavakoli, \href{https://doi.org/10.1103/PhysRevA.93.030101}   {Phys. Rev. A. {\bf 93}, 030101 (2016)}. 
	\bibitem{Frit2016}  T. Fritz, \href{
		https://doi.org/10.1007/s00220-015-2495-5}{Comm. Math. Phys. {\bf 341}, 391-434 (2016)}.
	\bibitem{Rosse2016}  D. Rosset, C. Branciard, T. J. Barnea, G. Pütz, N. Brunner,
	and N. Gisin, \href{https://doi.org/10.1103/PhysRevLett.116.010403}{Phys. Rev. Lett. {\bf 116}, 010403 (2016)}.
	\bibitem{Chav2016}  R. Chaves,  \href{https://doi.org/10.1103/PhysRevLett.116.010402} {Phys. Rev. Lett. {\bf 116}, 010402 (2016)}.
	\bibitem{Tava2017}  A. Tavakoli, M-O. Renou, N. Gisin, and N. Brunner, \href{https://iopscience.iop.org/article/10.1088/1367-2630/aa7673/pdf} {New J. Phys. {\bf 19}, 073003 (2017)}.
	
	\bibitem{Luo2018}  M-X. Luo, \href{https://doi.org/10.1103/PhysRevLett.120.140402}{Phys. Rev. Lett. {\bf 120}, 140402
		(2018)}.
	
	\bibitem{Gupta2018} S. Gupta, S. Dutta, and A. S. Majumder,  \href{https://journals.aps.org/pra/pdf/10.1103/PhysRevA.98.042322}  {Phys. Rev. A. {\bf98}, 042322 (2018)}. 
	\bibitem{Wolfe2019} E. Wolfe, R. W. Spekkens, and T. Fritz, \href{https://doi.org/10.1515/jci-2017-0020} {Jour. Causal Inference
		7, 2 (2019)}.
	\bibitem{Cyril2019} C. Branciard,  \href{
		https://doi.org/10.1038/s41566-019-0522-3}  {Nat. Photonics {\bf13}, 660–663  (2019)}.
	
	\bibitem{Kerstjens2019}A. Pozas-Kerstjens, R. Rabelo, \L. Rudnicki, R. Chaves, D. Cavalcanti, M. Navascues, and A. Acin, \href{https://doi.org/10.1103/PhysRevLett.123.140503} {Phys. Rev. Lett. {\bf 123}, 140503 (2019)}.
	\bibitem{Aberg2020} J. Åberg, R. Nery, C. Duarte, and R. Chaves, \href{https://doi.org/10.1103/PhysRevLett.125.110505}  {Phys. Rev. Lett. {\bf 125}, 110505  (2020)}.
	\bibitem{Banerjee2020} R. Banerjee, S. Ghosh, S. Mal, and A. Sen(De),
	\href{http://doi.org.10.1103/PhysRevResearch.2.043355}  {Phys. Rev. Research. {\bf 2},043355  (2020)}.
	\bibitem{Mukh2016} K. Mukherjee, B. Paul, and D. Sarkar,
	\href{DOI 10.1007/s11128-016-1301-4}{Quantum Inf. Process. {\bf15}, 2895 (2016)}
	\bibitem{Tava2016} A. Tavakoli,
	\href{10.1088/1751-8113/49/14/145304}{New J. Phys. A: Math. Theor. {\bf49} 145304 (2016)}
	\bibitem{Tavakoliarxiv} A. Tavakoli, N. Gisin, and C. Branciard,   \href{https://doi.org/10.1103/PhysRevLett.126.220401}  {Phys. Rev. Lett. {\bf126}, 220401 (2021) }.

	\bibitem{Gisinarxiv2021} J. D. Bancal, N. Gisin, \href{https://doi.org/10.1103/PhysRevA.104.052212} {Phys. Rev. A {\bf104}, 052212 (2021)}.
	\bibitem{Kraftarxiv} T. Kraft, S. Designolle,  C. Ritz,  N. Brunner,  O. Guhne, and  M. Huber, \href{https://doi.org/10.1103/PhysRevA.103.L060401} {Phys. Rev. A {\bf103}, L060401 (2021)}.
	\bibitem{kundu2020} A.Kundu, M. K. Molla, I. Chattopadhyay, and D. Sarkar, \href{https://doi.org/10.1103/PhysRevA.102.052222}  {Phys. Rev. A {\bf 102}, 052222 (2020)}.
	\bibitem{Andr2017}  F. Andreoli, G. Carvacho, L. Santodonato, R. Chaves, and F.
	Sciarrino, \href{https://iopscience.iop.org/article/10.1088/1367-2630/aa8b9b/pdf}{New J. Phys. {\bf 19}, 113020 (2017)}.
	\bibitem{munshi2021}  S. Munshi, R. Kumar, and A. K. Pan, 
	\href{https://doi.org/10.1103/PhysRevA.104.042217}{Phys. Rev. A {\bf104}, 042217 (2021).}
	\bibitem{munshi2021a}  S. Munshi and A. K. Pan, \href{https://doi.org/10.1103/PhysRevA.105.032216} {Phys. Rev. A, 105, 032216 (2022).}
	\bibitem{Carvacho2017} G. Carvacho, F. Andreoli, L. Santodonato, M. Bentivegna, R. Chaves, and  F. Sciarrino,   \href{ https://doi.org/10.1038/ncomms14775 }  {Nat Commun {\bf 8},14775 (2017)}.
	\bibitem{Saunders2017} D. J. Saunders, A. J. Bennet, C.  Branciard, and G. J. Pryde, \href{https://advances.sciencemag.org/content/3/4/e1602743}  {Sci. Adv.  {\bf 3},e1602743 (2017)}.
	\bibitem{Andreoli2017} F. Andreoli, G. Carvacho, L. Santodonato,  M. Bentivegna, R. Chaves, and  F. Sciarrino, \href{http://doi.org/10.1103/PhysRevA.95.062315} {Phys. Rev. A {\bf95}, 062315 (2017)}.
	
	\bibitem{Poderini2020} D. Poderini, \emph{et al.},  \href{https://doi.org/10.1038/s41467-020-16189-6} {Nat Commun {\bf 11}, 2467 (2020)}.
	\bibitem{Renou2019} M-O. Renou, E. Bäumer, S. Boreiri, N. Brunner, N. Gisin, and S. Beigi, \href{https://doi.org/10.1103/PhysRevLett.123.140401}{Phys. Rev.    Lett. {\bf 123}, 140401 (2019)}.
	\bibitem{Renou2022} M-O. Renou and S. Beigi,
	\href{https://doi.org/10.1103/PhysRevA.105.022408}{Phys. Rev. A {\bf105}, 022408 (2022) }
	\bibitem{Reno2022} M-O Renou and S. Beigi,
	\href{https://doi.org/10.1103/PhysRevLett.128.060401}{Phys. Rev. Lett. {\bf128}, 060401 (2022) }
	\bibitem{Gisi2020}  N. Gisin, \textit {et al.},   \href{https://www.nature.com/articles/s41467-020-16137-4.pdf} {Nat Commun {\bf 11}, 2378 (2020)}.
	\bibitem{Supic2020} I. Supic, J. Bancal, and N. Brunner,  \href{https://doi.org/10.1103/PhysRevLett.125.240403}  {Phys. Rev. Lett. {\bf 125 }, 240403 (2020)}.

\bibitem{Lee2018} C. M. Lee, and M. J. Hoban , \href{https://doi.org/10.1103/PhysRevLett.120.020504} {Phys. Rev.Lett. {\bf 120}, 020504 (2018)}.
	\bibitem{Kerst2022} A. P. Kerstjens, N. Gisin, and A. Tavakoli,
	\href{https://doi.org/10.1103/PhysRevLett.128.010403}{Phys. Rev. Lett. {\bf128}, 010403 (2022) }
	\bibitem{Supic2022} I. Supic, J-D. Bancal, Y. Cai, and N. Brunner,
	\href{https://doi.org/10.1103/PhysRevA.105.022206}{Phys. Rev. A {\bf105}, 022206 (2022)}
	\bibitem{Tava2021} A. Tavakoli , N. Gisin  and C. Branciard,
	\href{https://doi.org/10.1103/PhysRevLett.126.220401}{Phys. Rev. Lett. {\bf126}, 220401 (2021)}
		\bibitem{Ghorai2018} S. Ghorai, and A. K. Pan, 
	\href{https://doi.org/10.1103/PhysRevA.98.032110}{Phys.
		Rev. A {\bf 98}, 032110(2018)}.
	\bibitem{AKP2020} A. K. Pan, and S. S. Mahato, \href{https://doi.org/10.1103/PhysRevA.102.052221}{Phys.
		Rev. A {\bf 102}, 052221(2020)}.
	\bibitem{Asmita} A.Kumari, and A.K.Pan, \href{https://doi.org/10.1103/PhysRevA.100.062130}{Phys.
		Rev. A {\bf 100}, 062130(2019)}.

	
	
\end{thebibliography}
\end{document}